\documentclass[aps,twocolumn,groupedaddress,showpacs]{revtex4}
\usepackage[dvips]{graphicx}
\usepackage[]{caption}
\usepackage{amsmath}
\usepackage{amssymb}
\def\Dsl{\hbox{/\kern-.6700em\it D}} 
\def\dsl{\hbox{/\kern-.5300em$\partial$}}

\def\eqa{\begin{eqnarray}}
\def\eeqa{\end{eqnarray}}
\def\eq{\begin{equation}}
\def\eeq{\end{equation}}
\def\be{\begin{equation}}
\def\ee{\end{equation}}
\def\bea{\begin{eqnarray}}
\def\eea{\end{eqnarray}}

\begin{document}
\bibliographystyle{prsty}
\title{The Confining Heterotic Brane Gas: A Non-Inflationary Solution to the
  Entropy and Horizon Problems of Standard Cosmology}
\author{Robert Brandenberger and Natalia Shuhmaher}
\affiliation{Dept. of Physics, McGill University, 3600 University Street,
Montr\'eal QC, H3A 2T8, Canada}
\date{\today}
\begin{abstract}
We propose a mechanism for solving the horizon and entropy problems of
standard cosmology which does not make use of cosmological
inflation. Crucial ingredients of our scenario are brane gases, extra
dimensions, and a confining potential due to string gas effects which
becomes dominant at string-scale brane separations. The initial conditions
are taken to be a statistically homogeneous and isotropic hot brane gas in 
a space in which all spatial dimensions 
are of string scale. The extra dimensions which end up as the internal 
ones are orbifolded. The hot brane gas leads to an initial phase 
(Phase 1) of
isotropic expansion. Once the bulk energy density has decreased
sufficiently, a weak confining potential between the two orbifold fixed
planes begins to dominate, leading to a contraction of the extra spatial
dimensions (Phase 2). String modes which contain momentum about the 
dimensions perpendicular to the orbifold fixed planes provide a repulsive
potential which prevents the two orbifold fixed planes from colliding. The
radii of the extra dimensions stabilize, and thereafter our three spatial 
dimensions expand as in standard cosmology. The energy density after
the stabilization of the extra dimensions is of string scale, whereas the
spatial volume has greatly increased during Phases 1 and 2, thus leading 
to a non-inflationary solution of the horizon and entropy problems.  
\end{abstract}
\maketitle

\section{Introduction}

The Inflationary Universe scenario \cite{Guth} (see also
\cite{Sato,Brout,Starob})  has been extremely successful
phenomenologically. It has provided a solution to some of the key
problems of standard cosmology, namely the horizon and flatness
problems, and yielded a mechanism for producing
primordial cosmological perturbations using causal physics, a
mechanism which predicted \cite{ChibMukh,Lukash} (see also
\cite{Press,Sato}) an almost scale-invariant spectrum of adiabatic
cosmological fluctuations, a prediction confirmed more than a decade
later to high precision by cosmic microwave background anisotropy experiments
\cite{COBE,Boomerang,Maxima,WMAP}. 

In this paper, we will pay special attention to the ``entropy problem''
of standard cosmology \cite{Guth}. 
The problem consists of the fact that without accelerated
expansion of space, it is not possible to explain the large entropy, size and
age of our current universe without assuming that at very early times
the universe was many orders of magnitude larger than would be
expected on dimensional arguments. 

In the inflationary scenario, the entropy problem is solved by postulating a
sufficiently long period of accelerated expansion, after which the
universe reheats to a temperature comparable to that prior to the onset
of the period of acceleration. In most] models of inflation, the
accelerated expansion of space is sourced by the potential energy of a
slowly rolling scalar field. Such models, however, are subject to
serious conceptual problems (see e.g. \cite{RHBrev5,RHBrev1} for
recent overviews of these problems). Most importantly, the source of
the acceleration is very closely related to the source of the
cosmological constant in field theory, a constant which is between 60
and 120 orders of magnitude larger than the maximal value of the
cosmological constant allowed by current observations. Because of the
existence of these conceptual problems, it
is of great importance to look for possible alternatives to scalar
field-driven inflationary cosmology.

There have been various suggestions for alternative cosmologies. 
In varying speed of light models \cite{Moffatt,Joao}, postulating the
existence of a period in the early universe during which the speed of
light decreased very fast leads to a solution of the horizon problem. In
the ``Pre-Big-Bang scenario'' \cite{PBB}, the Universe is born cold, flat
and large, undergoes a period of super-exponential contraction before 
emerging into the period of radiation-dominated expansion of standard
cosmology. The contracting phase and the expanding phase are related
via a duality of string theory, namely ``scale-factor duality''. In
a more recent cosmological scenario motivated by heterotic M-theory
\cite{HW}, namely the ``Ekpyrotic scenario'' \cite{Ekp}, the collision
of a bulk brane onto our boundary orbifold fixed plane generates a 
non-singular expansion of our brane. However, neither the Pre-Big-Bang
nor the original Ekpyrotic scenario can explain why our Universe is so
large and old (without assuming that the Universe is already much
larger than would be expected by dimensional arguments at the end of
the phase of contraction (see e.g. \cite{PBBcrit,Pyro}) (this problem
is avoided in the ``cyclic scenario'' \cite{cyclic}, a further
development of ideas underlying the Ekpyrotic scenario, but this is
achived at the cost of additional ad hoc assumptions about the
cosmological bounce).  The size
problem has so far also prevented the ``string gas cosmology''
scenario \cite{BV,ABE} (see e.g. \cite{RHBrev2,WatBatrev} for recent
reviews) from making contact with late time cosmology, although
a stringy mechanism for producing a scale-invariant spectrum of
cosmological perturbations does exist in this context \cite{Ali1}.
  
In this paper, we present a potential solution of the entropy problem
which does not make use of a period of accelerated expansion. Our
solution makes use of several ingredients from string theory: extra
spatial dimensions, the existence of branes and orbifold fixed planes
as fundamental extended objects in the theory, and a stringy mechanism
for stabilizing the shape and volume moduli of string theory via the
production of massless string states at enhanced symmetry points in
moduli space. Thus, it is possible that our mechanism will find a
natural realization in string theory.

\section{Overview of the Model}

Our starting point is a topology of space in which all but three
spatial dimensions are orbifolded, and the three dimensions
corresponding to our presently observed space are
toroidal. Specifically, the space-time manifold is
\be
{\cal M} \, = \, {\cal R} \times T^3 \times T^d / Z_2 
\, ,
\ee
where $T^3$ stands for the three-dimensional torus, and $d$ is the
number of extra spatial dimensions, which we will take to be
either $d = 6$ in the case of models coming from superstring theory,
or $d = 7$ in the case of models motivated by M-theory. We will assume
that there is a weak confining force between the orbifold fixed planes
\footnote{It may be necessary to have branes pinned to the orbifold fixed
planes in order to induce such a potential. Our approach, at this stage,
is purely phenomenological, and we simply postulate the existence of a 
potential with the required properties}. 

As our initial conditions, we take the bulk to be filled with an
isotropic \footnote{Note that the orbifolding will prohibit the
existence of certain branes along certain of the dimensions and will
thus lead to a breaking of the condition of isotropy. The details
are fairly model-specific and will be discussed in a followup paper.
The bottom line, however, is that the noninflationary bulk expansion
of the first phase in all directions remains a valid conclusion.} gas of
branes, as in the studies of \cite{Mahbub,Mairi,Lisa}. These studies
show that, in the context of Type IIB superstring theory, the bulk of the 
energy density will end up in three and
possibly seven branes. However, if the initial Hubble radius is large
relative to the size of space, there will be no residual seven branes.
In the case of heterotic string theory or taking the starting point to
be M-theory, we would be dealing with Neveu-Schwarz 5-branes.  

Assuming that the universe starts out small and hot, it is reasonable
to assume that the energy density in the brane gas will initially be
many orders of magnitude larger than the potential energy density
generated by the force between the orbifold fixed planes. Thus,
initially our universe will be expanding isotropically. We denote this
as Phase 1. Our key observation is that in this phase, the energy
density projected onto the orbifold fixed planes does not decrease.
The reason is that the tension energy of the p-branes increases as
$a(t)^p$, where here $a(t)$ is the bulk scale factor. The volume
parallel to the orbifold fixed planes is increasing as $a(t)^3$,
and hence the projected energy density does not decrease (it is in fact
constant in the case of 3-branes).

During Phase 1, the bulk energy density will decrease. Hence,
eventually the inter-orbifold potential will begin to dominate. At
this point, the cosmological evolution will cease to be isotropic: the
directions parallel to the orbifold fixed planes will continue to
expand while the perpendicular dimensions begin to contract. We denote
this phase as Phase 2.

Once the orbifold fixed planes reach a microscopic separation, a repulsive
potential due to string momentum modes becomes important (one example is
the production of massless
states at enhanced symmetry points \cite{Watson,Beauty}). The interplay
between this repulsive potential which dominates at small separations and
the attractive potential which dominates at large distances, coupled to
the expansion of the three dimensions parallel to the orbifold fixed planes,
will lead to the stabilization of the radion modes at a specific radius 
(presumably related to the string scale). In the context of heterotic string
theory, we could use the string states which are massless at the
self-dual radius to obtain stabilization of the radion modes at the
self-dual radius \cite{Patil1,Patil2} (see also \cite{Watson2}). These
modes would also ensure dynamical shape moduli stabilization
\cite{Edna}.  
We denote the time of radion stabilization by $t_R$ since this time
plays a similar role to the time of reheating in inflationary
cosmology. The branes decay into radiation either during or at the
end of Phase 2. This brane decay is the main source of reheating of
our three dimensional space.

After the radion degrees of freedom have stabilized at a microscopic
value which presumably is set by the string scale, the three spatial
dimensions parallel to the orbifold fixed planes will continue to
expand. The energy density which determines the three-dimensional
Hubble expansion rate is the projected energy density $\rho_p$, i.e. the bulk
energy density integrated over the transverse directions. The key
point is that during Phase 1, $\rho_p$ does not decrease. If the
bulk is dominated by 3-branes, $\rho_p$ is constant, if it is
dominated by 5-branes, $\rho_p$ in fact increases. In Phase 2,
the projected energy density $\rho_p$ also remains constant if
the bulk is dominated by 3-branes, modulo
the conversion of brane tension energy into radiation as the bulk
branes decay or are absorbed by the fixed planes. If we approximate the
evolution by assuming that all of the brane tension energy converts to
radiation at the time of radion stabilization, then
the value of $\rho_p$ at $t_R$, which is the energy
density which determines the evolution of the scale factor of our
three spatial dimensions after $t_R$, is equal to the projected energy
density at the initial time, which we take to be given by the string scale
\footnote{We assume that initial radii and densities are all set by the
string scale, i.e. we introduce no unnaturally small or large numbers.}.
If the branes decay during Phase 2, then the projected energy density at
$t_R$ is larger than the initial value, in which case we may be driven
to a Hagedorn phase of string theory.  
The main point, however, it that since the volume of our three 
spatial dimensions has been
expanding throughout Phases 1 and 2, the horizon and entropy problems of
standard cosmology can easily be solved by simply assuming that the
phase of bulk expansion lasted sufficiently long (numbers will be
given later).

Note that we are assuming in this paper that the dilaton has been
stabilized by some as yet unknown mechanism. In this case, the
equations of motion of the bulk are those of homogeneous but
anisotropic general relativity. The metric is in this case given by
\be
ds^2 \, = \, dt^2 - a(t)^2 d{\bf x}^2 - b(t)^2 d{\bf y}^2 \, ,
\ee
where ${\bf x}$ denote the three coordinates parallel to the boundary
planes and ${\bf y}$ denote the coordinates of the perpendicular
directions. In the case of $d$ extra spatial directions the equations
of anisotropic cosmology are  
\bea \label{dyn1}
{\ddot a} &+& {\dot a} \bigl( 2H + d{\cal H} \bigr) \\
&=& \, 8 \pi G a \bigl[ P - {1 \over {3 + d - 1}} (3P + d{\tilde P}) +
  {1 \over {3 + d - 1}} \rho \bigr] \, , \nonumber 
\eea
\bea \label{dyn2}
{\ddot b} &+& {\dot b} \bigl( 3H + (d - 1){\cal H} \bigr) \\
&=& \, 8 \pi G b \bigl[ {\tilde P} - {1 \over {3 + d - 1}} (3P +
  d{\tilde P}) + {1 \over {3 + d - 1}} \rho \bigr] \, , \nonumber
\eea
and
\be \label{constr}
(3H + d{\cal H})^2 - 3H^2 - d{\cal H}^2 \, = \, 16 \pi G \rho \, ,
\ee
where $H \equiv {\dot a}/a$, ${\cal H} \equiv {\dot b}/b$ are the
expansion rates of the parallel and perpendicular dimensions,
respectively, $\rho$ is the bulk energy density and $P$ and ${\tilde P}$
are the parallel and perpendicular pressures, respectively.

\section{The Phase of Bulk Expansion}

During the phase of bulk expansion, the two scale factors coincide, $P
= {\tilde P}$, and both equations (\ref{dyn1}) and (\ref{dyn2}) reduce
to
\be \label{bulk}
{{{\ddot a}} \over a} + (2 + d)\bigl({{{\dot a}} \over a} \bigr)^2
\, = \, {{8 \pi G} \over {3 + d - 1}} \bigl[ \rho - P \bigr] \, .
\ee 
Making use of the equation of state $P = w \rho$, and inserting
(\ref{constr}), the dynamical equation (\ref{bulk}) becomes
\be \label{bulk2}
{{{\ddot a}} \over a} + (2 + d)\bigl({{{\dot a}} \over a} \bigr)^2
\, = \, {1 \over 2} (3 + d)(1 - w) \bigl({{{\dot a}}
  \over a} \bigr)^2 \, ,
\ee
which leads to power law expansion
\be \label{scaling2}
a(t) \, \sim t^{\alpha} \,
\ee
where the value of $\alpha$ depends on the equation of state:
\be \label{scaling3}
\alpha \, = \, {2 \over {(3 + d)(1 + w)}} \, .
\ee 
If the bulk energy is dominated by the tension of $p$-branes, then we have
\be
w \, = \, - {p \over {3 + d}} \, .
\ee
In the example motivated by perturbative Type IIB superstring theory,
namely $d = 6$ and $p = 3$ we obtain
\be \label{scaling1}
a(t) \, \sim \, t^{2 / (3 + d - p)} \, = \, t^{1/3} \, .
\ee
What is important for us is that this is not accelerated expansion.
Starting with heterotic string theory, we would have $d = 6$ and $p = 5$
and for M-theory we would take $d = 7$ and $p = 5$. These two cases
lead to faster expansion rates, namely $\alpha = 1/2$ in the former case
and $\alpha = 2/5$ in the latter. 

\section{The Phase of Orbifold Contraction}

If we want the expansion which takes place in this initial phase to 
solve the size and horizon problems of standard cosmology independent
of any further expansion during Phase 2, then the scale factor needs
to increase by a factor ${\cal F}$ of at least
\be \label{need}
{\cal F} \, \sim \, 10^{30} \, . 
\ee
This result comes about by demanding that the predicted radius of
the universe evaluated at the present temperature 
be greater than the presently observed Hubble radius, i.e. greater
than $10^{42} \rm{GeV}^{-1}$, by taking the density at the time $t_R$
to be given by the string scale which we take to be $10^{17}$GeV, and 
taking into account that the scale factor in standard cosmology
increases by a factor of about $10^{29}$ between when the temperature
is of string scale and today. Correspondingly, the radiation
temperature of the bulk will decrease by the same factor ${\cal F}$.

We now assume the existence of a confining potential
$V$ between the orbifold fixed planes. In order to generate
such a non-vanishing potential, we will need to assume that
branes are stuck to the orbifold fixed planes. In terms of the distance
$r = l_s b$ between these planes ($l_s$ being the string length), 
a typical confining potential is
\be \label{pot}
V(r) \, = \, \mu r^n \, = \, \mu (l_s b)^n ,
\ee
where $n$ is an integer, $\mu \equiv \Lambda^{d+n+4}$, and 
$\Lambda$ is the typical energy scale of the potential. As we will
show below, a value $n \, \geq \, 2 \sqrt{2d} + d$ 
is required for our scenario to work.

The presence of this potential will lead to a transition between the
phase of isotropic expansion to a phase in which the extra 
dimensions contract while the dimensions parallel to the fixed planes
keep on expanding (and we will verify below that the expansion is
not inflationary). The transition between Phase 1 and Phase 2 takes
place when the bulk energy density and the inter-brane potential
become comparable. The bulk energy density in Phase 1 scales as
\be
\rho_b(t) \, \sim \, b(t)^{- d - 3 + p} 
\ee
(recall that in this phase $a(t) = b(t)$). Assuming that the
initial bulk energy density is set by the string scale, and using the result
(\ref{scaling1}), it follows that in order for the bulk to have
expanded by the factor of (\ref{need}), the upper bound on $\Lambda$ 
should satisfy: 
\be \label{Lambda} 
\Lambda \, \sim \, l_s^{-1} 10^{- 30{{d - p + 3 + n} \over {d + 4 + n}}} \, .
\ee
In the case $d = 6$, $p = 3$, and $n = 2$ we obtain
\be
\Lambda \, \sim \, l_s^{-1} 10^{-20} \, \sim 1{\rm MeV} \, .
\ee
For $d = 6$ and $p = 5$ the result is 
$\Lambda \sim l_s^{-1} 10^{-15} \sim 100 {\rm GeV}$,
for $d = 7$ and $p = 5$ we obtain 
$\Lambda \sim l_s^{-1} 10^{-210/13} \sim 10 {\rm GeV}$. Note that 
these values for $\Lambda$ are not too different from the scale of 
electroweak symmetry breaking.  

We will analyse the evolution during Phase 2 using a four-dimensional
effective field theory, where we replace the radion $b(t)$ by a scalar
field $\varphi(t)$. In order that $\varphi$ be canonically normalized
when starting from the higher dimensional action of General Relativity,
$\varphi$ and $b$ must be related via (see e.g. \cite{WatBatrev},
Appendix A)
\be
\varphi \, = \, m_{pl} \sqrt{2d}{\rm log}(b) \, ,
\ee
where $m_{pl}$ is the four-dimensional Planck mass. If the bulk
size starts out at the string scale, then $b(t_i) = 1$, where $t_i$
is the initial time. With these normalizations, $\varphi = 0$
corresponds to string separation between the branes. In terms of
$\varphi$, the potential (\ref{pot}) then induces an effective
potential for $\varphi$:
\be
V_{eff}(\varphi) \, = \, \mu l_s^{(n + d)} e^{{\tilde n} \varphi / m_{pl}} \, ,
\ee
where ${\tilde n} = (n - d) / \sqrt{2d}$.
Note that the original bulk potential needs to be multiplied by the area
of the orbifold fixed plane in order to obtain the effective potential
for $\varphi$, $V_{eff}(\varphi)$. There is also a factor of $b^{-2d}$
coming from converting to the Einstein frame (see e.g. \cite{WatBatrev},
Appendix A). The equation of motion for $\varphi$ then becomes
\be \label{eom1}
{\ddot \varphi} + 3H {\dot \varphi} \, 
= \, - {\tilde n} {{\mu l_s^{(n + d)} \over {m_{pl}}}} 
e^{{\tilde n} \varphi / m_{pl}} \, 
\ee
with 
\be \label{eom2}
H^2 \, = \, {1 \over {3 {m_{pl}^2}}} \bigl[{{\dot \varphi}^2 \over 2} 
+ {{\mu l_s^{(n+d)}} \over {m_{pl}}} e^{{\tilde n} \varphi / m_{pl}}\bigr] \, 
\ee

During Phase 2, the scale factor $a(t)$ of the three spatial dimensions
parallel to the orbifold fixed planes will expand according to the
usual four space-time dimensional cosmological equations, where
matter is dominated by the scalar field $\varphi$. The solution of 
the equations of motion~(\ref{eom1} and~\ref{eom2}) in the cases 
${\tilde n} = 1$ and ${\tilde n} = 2$ is given by 
\be \label{solution}
\varphi \, = \, {m_{pl} \over {\tilde{n}}} 
\ln{{2 m^2_{pl}(6-{\tilde n}^2)} \over {{\tilde n}^4 \mu {l_s}^{(n + d)} t^2}} \, 
\ee
The corresponding values of the equation of state parameter are 
\be
w \, = \, {{\tilde n}^2-3 \over 3} \, .
\ee
For ${\tilde n} = 1$ this equation of state corresponds to an accelerating
background, but for ${\tilde n} = 2$ the background evolution is non-accelerating.
In fact, as ${\tilde n}$ grows one can easily show that the usual 
inflationary slow-roll conditions are grossly violated. Thus, for a value of 
${\tilde n} \geq 2$ or equivalently $n \, \geq \, 2 \sqrt{2d} + d$
the evolution of $a(t)$ during this phase will be non-inflationary.

Taking into account the bulk expansion during Phase 1 of (\ref{need}),
it follows that for $d = 6$ and $p = 3$
the initial value of $\varphi$ is about $69 m_{pl}$.
The exponential
form of the potential will lead to a rapid collapse of the extra
dimensions. To estimate the time scale of the decrease, we replace
the source of the right hand side of (\ref{eom1}) by its initial value
and estimate the time interval $\Delta t$ for $\varphi$ to decrease by
an amount $m_{pl}$. We find that this time interval equals the initial
Hubble time. Thus, a rough estimate of the duration of Period 2 is
$10^2 H^{-1}$.
 
\section{Modulus Stabilization and Late Time Cosmology}

The next crucial step in our scenario is to invoke a mechanism to
stabilize the radius of the extra dimensions at a fixed radius. Such
modulus stabilization mechanisms have recently been extensively studied both
in the context of string theory models of inflation (see e.g. 
\cite{stringinflation} for recent reviews)
and in string gas cosmology \cite{RHBrev3}. We will make use of the
mechanism developed in the latter approach.

String modes which carry momentum about the extra dimensions will generate
an effective potential for the radion which is repulsive. These
repulsive effects will dominate for values of the radion smaller than
the self-dual radius. Since these modes are very light at large values of
the radion, it is likely that they will be present in great abundance.
Even if they are not, the subset of such modes which are massless at
enhanced symmetry points will be copiously produced when the value
of the radion approaches such points \cite{Watson,Beauty}
\footnote{As discussed in \cite{Patil2}, stabilization via string modes
which are massless at the self-dual radius leads to a consistent late time
cosmology.}. The
induced potential will lead to a source term in the equation
of motion for the scale factor $b(t)$ which is of the form 
\cite{Patil1,Patil2}
\be
{\ddot b} + 3H {\dot b} \, = \, 8 \pi G n(t) 
\bigl[\bigl({1 \over b}\bigr)^2 - b^2 \bigr] + ... \, ,
\ee
where the dots indicate extra source terms from other string modes,
as well as terms quadratic in ${\dot b}$. Note that $n(t)$ is given
by the number density of the modes. Translating to the
scalar field $\varphi$, and neglecting terms quadratic in ${\dot \varphi}$,
the above equation becomes
\be
{\ddot \varphi} + 3H {\dot \varphi} \, = \, 8 \pi G n(t)
e^{-\varphi / m_{pl}} m_{pl} 
\bigl( e^{-2\varphi / m_{pl}} - e^{2\varphi / m_{pl}} \bigr) \, .
\ee
Thus, it follows that after approaching the self-dual radius, $b(t)$
will perform damped oscillations about $b(t) = 1$, or, in other words,
$\varphi(t)$ will undergo damped oscillations about and get
trapped at $\varphi = 0$ (which corresponds to string scale separation
between the orbifold fixed planes). At this separation, the four 
dimensional effective potential $V_{eff}$ becomes
\be
V_{eff} \, = \, \Lambda^{d + 4 + n} l_s^{n + d} \, ,
\ee
and, taking upper limit on $\Lambda$ from (\ref{Lambda}), this becomes 
\be
V_{eff} \, = \, l_s^{-4} 10^{-30(d-p+3+n)} \, .
\ee
Thus, starting with vanishing cosmological constant in the bare
bulk Lagrangian, our scenario accidentely generates a cosmological constant
energy density in our present universe which is suppressed by 
$30 \times (d-p+3+n)$ orders of magnitude. This will provide the
correct order of the cosmological constant to account for the current
acceleration if $d=p$ and $n=1$.  

Either at some point during the phase of contraction, or else when
the distance between the orbifold fixed planes has decreased to the
string scale, all of the bulk branes will decay, presumably predominantly
into radiation along the fixed plane directions.
The three unconfined spatial dimensions will thus
emerge in the expanding radiation-dominated phase of standard cosmology.
The energy density which at late time governs the dynamics of our
scale factor $a(t)$ is the bulk energy density integrated over the
transverse dimensions. Since the bulk energy in Phase 1 is dominated
by the $p = 3$ branes, the integrated energy density is constant.
Thus, at the beginning of the radiation-dominated phase the effective
energy density is of the same order of magnitude as the initial bulk energy 
density, namely given by the string scale.

From the point of view of late time cosmology, what has been achieved
during Phase 1 is to increase the size of our spatial sections 
without decreasing the effective energy density. Without extra spatial
dimensions, the energy density can only remain constant if the expansion
of space is inflationary, but making use of the dynamics of extra
spatial dimensions, constant effective energy density can be achieved
using non-accelerated expansion of all dimensions.

Note that in the case of $p > 3$, specifically in the cases where we
use Neveu-Schwarz 5-branes in the bulk, the projected energy density
actually increases in Phase 1. If it decreases less during Phase 2 than
it increased during Phase 1 (which will be the case e.g. if the branes
convert to radiation during Phase 2), then the possibility emerges that
we are driven to a Hagedorn phase of string theory towards the end of
Phase 2 \cite{BV,TV}. In this case, a very nice mechanism for the generation
of a scale-invariant spectrum of fluctuations \cite{Ali1} can be realized.
This possibility will be briefly discussed in the next section.

There is another key prediction of our model which is closely related
to the chosen topology of space. No odd-dimensional cycles exist on the 
inner space $T^6 / Z_2$, thus prohibiting 
certain stable configurations of p-branes. Given that we are using
odd-dimensional branes in our examples, only $1$ or $3$ brane dimensions can 
wrap our three-dimensional toroidal space $T^3$, because no odd-dimensional
stable p-branes can have an odd number of their brane dimensions wrapped 
about the inner space. This prevents the creation of stable ``stringy'' 
domain walls and monopoles in our universe, but it may predict the
existence and future detection of cosmic strings.

\section{Discussion and Conclusions}

By making use of some tools coming from string theory, we have
proposed a mechanism to solve the entropy (size) problem of standard cosmology
without inflation. According to our proposal, the universe begins
hot, small and dense. We assume that the six extra spatial dimensions
of perturbative superstring theory are orbifolded, the three
dimensions we see today are not (they are toroidal). The universe
emerges with a gas of bulk branes (e.g. three branes if we have the
perturbative limit of Type IIB superstring theory in mind or 5-branes
if we start from heterotic string theory or M-theory)
which drives an initial phase of isotropic bulk expansion of all nine
spatial dimensions. During this phase, the energy density projected
onto the orbifold fixed planes does not decrease, even though the scale
factor is expanding (as $t^{1/3}$ in the case of 3-branes
in six extra dimensions). We assume the presence of a weak confining
potential between the orbifold fixed planes (the cosmological scenario
which emerges when considering a more conventional type of potential
will be discussed in a followup paper \cite{Natalia2}). Such a potential
will eventually dominate over the bulk energy density and will lead
to a second phase in which the extra spatial dimensions rapidly
contract while our three spatial dimensions continue to expand.
Once the orbifold fixed planes approach each other to within the
string scale, stringy effects previously studied in the context
of string gas cosmology will stabilize the radion degrees of freedom.
The bulk branes decay, and the universe emerges into the radiation-dominated
phase of standard cosmology, with a temperature which is of string scale,
but a size which is many orders of magnitude larger than what would
be expected on dimensional arguments \footnote{Note that our proposal
has certain similarities with the approach of \cite{Trodden}, in which
- in the context of brane world cosmology -
it was proposed that the decay of Kaluza-Klein bulk modes will lead
to an entropy flow from the bulk to the brane which can solve the
entropy and homogeneity problem of standard cosmology without
requiring a phase of inflationary expansion.}. Note that in the proposed 
scenario, the electroweak scale and the late time acceleration may be 
explained by the same source.

Since the initial spatial
section is in thermal contact, the horizon problem of standard cosmology
is explained, as well. Our scenario, however, does not solve the
flatness problem of standard cosmology. If the initial spatial sections
are curved, then the curvature will lead to a re-collapse of the
universe. One way to address the flatness problem is to invoke a special
symmetry such as the BPS symmetry (see e.g. \cite{BPS} for
a textbook discussion) which prohibits spatial curvature.

In order to provide an alternative to inflation in terms of solving
all of the cosmological problems of standard cosmology which inflation 
addresses, we need to find a mechanism for generating fluctuations.
Work on this topic is in progress. Since the universe is initially in
causal contact, there are no causality arguments which prevent the
generation of adiabatic fluctuations. It is possible that bulk
fluctuations similar to the ones proposed in the Ekpyrotic scenario
could play this role. Provided there are scale-invariant fluctuations
in bulk metric variables during the contracting phase, the work of
\cite{Thorsten} (see also \cite{Tolley}) shows that such fluctuations
will induce a scale-invariant spectrum of four dimensional metric
fluctuations in the radiation-dominated phase. Another possibility,
in particular in the context of branes with spatial dimension larger
than three, is that the post-collapse phase will lead to such high
densities that a quasi-static Hagedorn phase will result. The
Hagedorn phase makes a smooth transition to the radiation-dominated
phase of standard cosmology. In this
case, string thermodynamics automatically generates a scale-invariant
spectrum of adiabatic fluctuations on all scales smaller than the
Hubble radius during the quasi-static phase \cite{Ali1}. 

\begin{acknowledgments}

We wish to thank Keshav Dasgupta for useful discussions and for
comments on our manuscript, and Thorsten Battefeld for pointing
out a scaling error in an initial draft of the paper.
This work is supported by funds from McGill University,
by an NSERC Discovery Grant and by the Canada Research Chairs program.
N.S. would like to acknowledge support from a Carl Reinhardt
McGill Major Fellowship.

\end{acknowledgments}

\end{document}